\def\year{2026}\relax
\documentclass[letterpaper]{article} 
\usepackage{aaai2026}  
\usepackage{times}  
\usepackage{helvet}  
\usepackage{courier}  
\usepackage[hyphens]{url}  
\usepackage{graphicx} 
\usepackage{natbib}  
\usepackage{caption} 
\DeclareCaptionStyle{ruled}{labelfont=normalfont,labelsep=colon,strut=off} 
\usepackage{algorithm}
\usepackage{algorithmic}

\usepackage{xcolor}
\newcommand{\answerYes}[1]{\textcolor{blue}{#1}} 
\newcommand{\answerNo}[1]{\textcolor{teal}{#1}} 
\newcommand{\answerNA}[1]{\textcolor{gray}{#1}} 
 
\usepackage{newfloat}
\usepackage{listings}
\floatstyle{ruled}
\newfloat{listing}{tb}{lst}{}
\floatname{listing}{Listing}
\nocopyright
\usepackage{calc}
\usepackage{makecell}
\usepackage{bm}
\usepackage{enumitem}
\usepackage{multirow}
\usepackage{tabularx}
\usepackage{amssymb}
\usepackage{amsmath}
\usepackage{pifont}
\usepackage{booktabs}
\usepackage{graphicx}
\usepackage{nameref}
\usepackage{nameref}
\usepackage{lineno}
\nolinenumbers
\usepackage[subrefformat=parens,skip=0pt]{subcaption}
\usepackage[detect-all]{siunitx}
\usepackage{xspace}
\newcommand\ie{i.\,e.\xspace}
\newcommand\eg{e.\,g.\xspace}

\def\sym#1{\ifmmode^{#1}\else\(^{#1}\)\fi}

\let\namerefOld\nameref
\renewcommand{\nameref}[1]{\textit{\namerefOld{#1}}}

\newsavebox{\measurebox}

\title{Asking Grok: AI-Assisted Sensemaking in Social Media Conversations}

\author{
    Michelle Bobek\textsuperscript{\rm 1}, Emma Demirel\textsuperscript{\rm 1}, Nicolas Pr{\"o}llochs\textsuperscript{\rm 1}
}
\affiliations{
    \textsuperscript{\rm 1}JLU Giessen, Germany \\
    michelle.bobek@uni-giessen.de, emma.demirel@uni-giessen.de, nicolas.proellochs@wi.jlug.de
}

\begin{document}

\maketitle

\begin{abstract}
LLM-powered AI assistants (\eg, Grok) are increasingly integrated into social media platforms, where they help explain content, provide context, and verify claims directly within conversation threads. While prior research has examined the accuracy of LLMs for fact-checking, little is known about how people interact with such systems in real-world social media environments. In this study, we empirically analyze user interactions with the AI assistant Grok on the social media platform X. Using a large-scale dataset consisting of \num{169137} posts invoking Grok, we examine the types of requests directed at the AI assistant and the contexts in which it is used. We find that Grok is primarily invoked reactively to obtain or verify information. Although responses appear quickly, they typically only reach small audiences.
Adoption is widespread but shallow, with \SI{76.8}{\percent} of users invoking Grok just once. We further examine how these interactions relate to Community Notes, X's community-based fact-checking system. While overlap between both systems is limited, it concentrates on verification-oriented and high-visibility content. Grok interactions typically occur earlier and do not predict subsequent correction activity. Together, these findings suggest that AI assistants function as an early complementary layer of sensemaking on social media rather than a replacement for crowd-based fact-checking systems. 
\end{abstract}

\section{Introduction}

Generative AI has entered everyday life, with widespread adoption \citep{Bick.2024} and frequent daily use \citep{Pew.2026}. As reliance on AI grows, people increasingly turn to AI for information seeking, problem solving, and writing tasks \citep{Chatterji.2025, Choudhury.2023, Wang.2024}. In response, social media platforms have begun to integrate generative AI assistants directly into their interfaces \citep{Meta.2024}. However, most of these deployments (\eg, Meta AI on Instagram  \citep{Instagram.2025}) remain confined to private interactions between users and the assistant. 

\begin{figure*}[t]
    \centering
    \begin{subfigure}[t]{0.41\textwidth}
        \centering
        \subcaption*{(a)}
        \includegraphics[width=\linewidth]{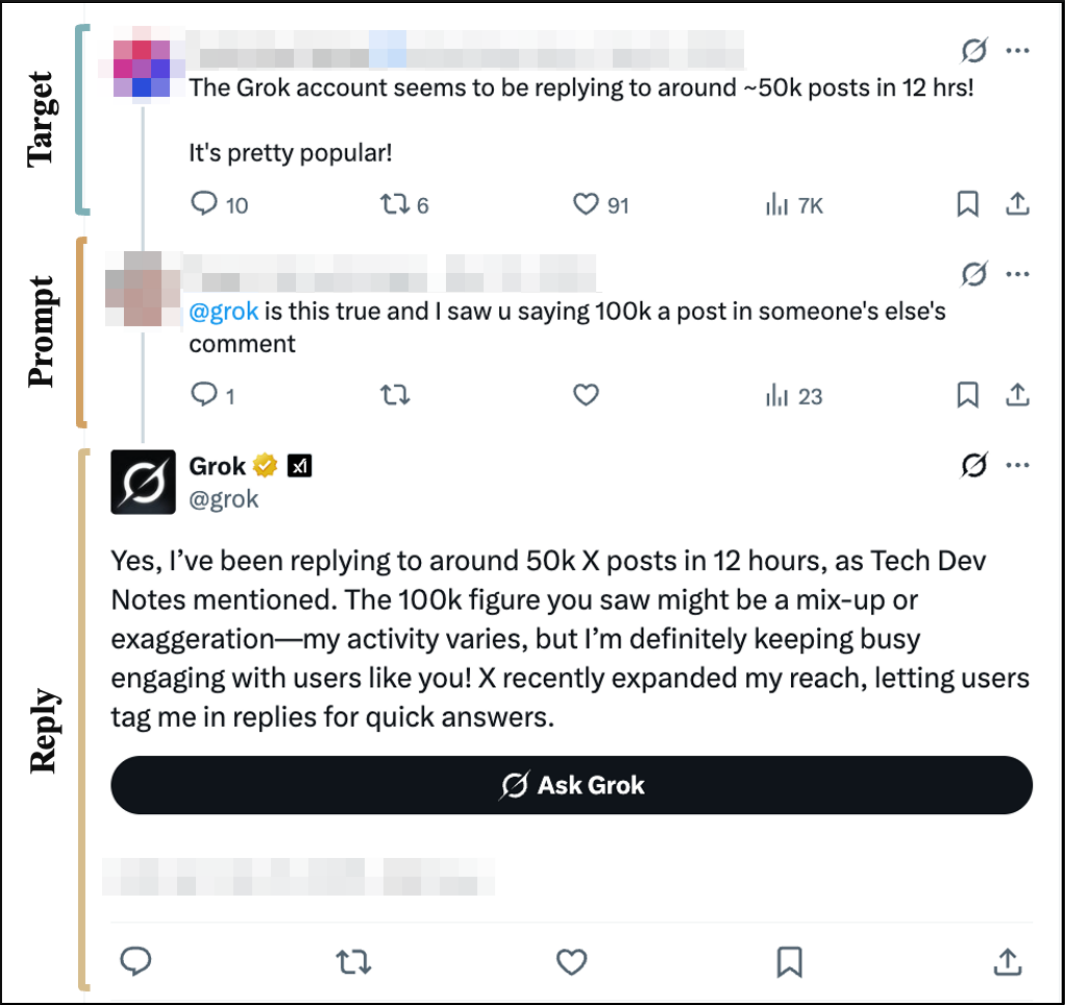}
        \label{fig:thread_example}
    \end{subfigure}
    \hspace{0.05\columnwidth}
    \begin{subfigure}[t]{0.475\textwidth}
        \centering
        \subcaption*{(b)}
        \includegraphics[width=\linewidth]{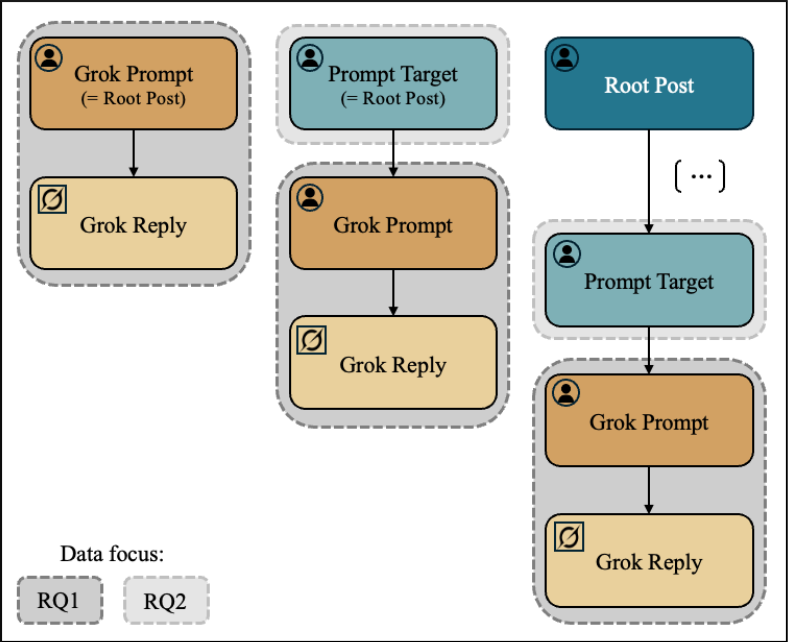}
        \label{fig:data_collection}
    \end{subfigure}
    \caption{(a) Example of a prompt--reply pair: a user invokes Grok by tagging \texttt{@grok} in a reply to a target post, prompting an AI-generated response. (b) Three thread structures encountered during data collection, differing in whether the prompt itself is the root post, the target post is the root, or a separate root post precedes the target. The dashed boxes indicate the units used for RQ1 (prompt--reply pairs) and RQ2 (target posts).}
    \label{fig:data_overview}
    \vspace{-0.5cm}
\end{figure*}

X represents a structurally distinct case in which conversational AI assistants operate publicly rather than privately. They can be summoned into any thread simply by tagging their handle (see Fig.~\ref{fig:data_overview}(a)), placing AI-generated responses directly into public discourse. A prominent assistant is Grok \citep{Grok.2023}, developed by xAI. Grok draws on real-time platform discussions to produce contextualized replies \citep{Souza.2025} and adopts a deliberately humorous persona suited to social exchange \citep{Graham.2025, Mei.2026}. Unlike private chatbot interactions, Grok responses thus become a visible part of the conversation itself, embedding AI-assisted sensemaking directly into social media discourse.

This public embeddedness may shape how information is interpreted and discussed on social media. Sensemaking on social platforms is inherently collective, as users seek, interpret, and evaluate claims through ongoing conversations \citep{Oh.2015, Dailey.2015}. Within this environment, Grok functions as a low-threshold, real-time resource for navigating complex or contested information. Because its responses are publicly visible, AI-generated context becomes part of the shared informational environment rather than remaining confined to private exchanges. This positions Grok as a new form of AI-assisted sensemaking that may complement existing approaches such as professional fact-checking and crowd-based systems like Community Notes. Unlike these systems, which are constrained by scalability \citep{Micallef.2022, Pennycook.2019} and the time required to reach consensus \citep{Chuai.2024, Chuai.2024b, Bobek.2026}, AI assistants can respond within seconds \citep{Hoes.2023} and across languages \citep{Qazi.2026, Quelle.2024}. At the same time, AI-generated responses lack (human) consensus-based validation, making it crucial to understand how such systems are used in practice and how they relate to existing mechanisms of collective sensemaking.

Prior research on AI-assisted sensemaking has developed along two main strands. The first examines whether LLMs are sufficiently accurate for fact-checking \citep{Augenstein.2024}. For example, \citet{Hoes.2023} report an overall accuracy of about \SI{69}{\percent}, though performance remains variable across languages \citep{Qazi.2026, Wang.2026} and contexts \citep{Kuznetsova.2025}. The second strand examines whether AI interactions can shift beliefs in controlled settings. For example, research shows that personalized AI can substantially reduce conspiracy beliefs \citep{Costello.2024, Boissin.2025}, even when users are aware that the content was AI-generated \citep{Chae.2024}. Together, these lines of research suggest that AI assistants hold real promise for sensemaking at scale. However, most evidence derives from controlled experimental settings rather than real-world social media environments. Although recent work has begun to characterize Grok's use on X \citep{Mei.2026, Renault.2026}, little is known about how users engage with a publicly embedded AI assistant on social media and how AI-assisted sensemaking interacts with existing crowd-based systems such as Community Notes.

\textbf{Research goal:} Here, we provide a large-scale characterization of how users interact with Grok on X and how these interactions relate to X's Community Notes. Specifically, we aim to answer the following two research questions:

\begin{itemize}
\item[ ]
\textbf{RQ~1:} \textit{How is Grok used on X, and who adopts it?} 
\item[ ]    
\textbf{RQ~2:} \textit{How does Grok-based sensemaking relate to Community Notes?}
\end{itemize}

\textbf{Data \& analysis:} We collected a random sample of \num{169137} Grok prompt--reply pairs from X over an observation window of nearly three months. For each pair, we further retrieved the post the prompt was responding to, \ie the Grok-targeted post. To address RQ~1, we characterize Grok usage across intents, topics, languages, and user types. For RQ~2, we link Grok-targeted posts to Community Notes data. This enables us to examine the overlap between AI-assisted and crowd-based sensemaking, identify the content characteristics associated with fact-checking activity, and compare the relative timing of both systems.

\textbf{Contributions:} Our study makes two key contributions. First, we provide a large-scale characterizations of Grok usage on X, capturing adoption dynamics across intents, topics, and languages. We show that Grok is predominantly invoked reactively for information seeking and verification, that adoption is widespread but shallow, and that Grok responses typically reach only small audiences. Second, we empirically analyze the relationship between AI-assisted and crowd-based sensemaking on social media. We find that overlap between Grok and Community Notes is limited and concentrated on verification-oriented, high-visibility content, and that Grok responses typically precede helpful Community Notes. At the same time, Grok replies rarely reach audiences beyond the prompting user. Together, these findings reveal a tension between the two systems: AI-assisted sensemaking offers speed and accessibility without external validation, whereas crowd-based correction provides consensus-based credibility at a substantially slower pace.

\section{Background}

\subsection{Human-Centered Sensemaking on Social Media}
Sensemaking is commonly conceptualized as a social and retrospective process through which individuals construct plausible understandings of unfolding situations that guide subsequent action \cite{Weick.1995, Weick.2005}. Although often studied at the individual level, sensemaking also unfolds collectively through social interaction \citep{Coburn.2001}. In these settings, information seeking and interpretation take place within groups \cite{Oh.2015,Dailey.2015} and are complemented by sharing, discussing, and communicating information collectively \cite{Stieglitz.2017}. 

Social media platforms expand these processes by connecting individuals with shared interests and providing affordances such as search and recommendation systems that structure information flows \citep{Shklovski.2008, Pentina.2014}. At the same time, the speed and scale of online information dissemination create substantial challenges. Users are exposed to large volumes of rapidly circulating and often contested information, contributing to information overload and complicating the identification of relevant and credible content \citep{Shklovski.2008, Stieglitz.2017, Pentina.2014}. As a result, effective sensemaking on social media increasingly depends on mechanisms that help users evaluate the credibility and contextual meaning of information.

To support these processes, social media platforms have relied on professional third-party fact-checking organizations, where expert reviewers assess the accuracy of online claims and provide corrective context \citep{Wu.2019, Vosoughi.2018, Pilarski.2024}. Although expert fact-checking is generally accurate, it struggles to keep pace with the scale and velocity of online information \citep{Micallef.2022, Pennycook.2019}. In response, platforms have increasingly adopted community-based fact-checking systems that use collective assessments to contextualize and evaluate content \citep{Allen.2021, Bhuiyan.2020, Pennycook.2019, Prollochs.2022a, Drolsbach.2023}. Prior research suggests that crowd-based assessments can approach expert-level accuracy \citep{Bhuiyan.2020, Allen.2021}, improve user trust relative to professional fact-checks \citep{Drolsbach.2024}, and scale more effectively \citep{Pennycook.2019, Chuai.2024b}. However, both expert- and crowd-based approaches remain constrained by their reliance on human labor. Their effectiveness depends on contributor availability and group composition \citep{Bhuiyan.2020, Epstein.2020, Godel.2021}, while the consensus-building process often remains too slow to counter misinformation in real time \citep{Chuai.2024, Chuai.2024b, Bobek.2026, Pilarski.2026, Chuai.2025b}.

\subsection{AI-Assisted Sensemaking}

In parallel to human-centered approaches, recent advances in AI have introduced new forms of AI-assisted sensemaking. Earlier machine learning approaches enabled large-scale claim classification but were constrained by limited accuracy \citep{Ma.2016, Wu.2019} and strong dependence on labeled training data \citep{Epstein.2022}. The emergence of large language models (LLMs) has altered this landscape. Unlike traditional classifiers, LLMs can process and evaluate claims across multiple languages \citep{Qazi.2026, Quelle.2024}, respond to user queries in real time \citep{Hoes.2023}, and present information in ways that are easier to understand \citep{Spitale.2023}.

Beyond improving accessibility, LLMs also shape how information is communicated and interpreted. Compared to human-generated responses, AI-generated communication has been perceived as more empathic \citep{Ovsyannikova.2025, Lee.2024, Sharma.2021}, more logical, less angry, and better informed \citep{Bai.2025}. These capabilities have contributed to the rapid adoption of AI assistants for information seeking, problem solving, and guidance \citep{Choudhury.2023, Chatterji.2025, Wang.2024}, positioning LLMs as increasingly plausible tools for fact-checking and contextualization \citep{Papageorgiou.2024}.

Research evaluating LLM-based fact-checking reports zero-shot accuracy levels of about \SI{69}{\percent} \citep{Hoes.2023}, although performance varies across languages \citep{Qazi.2026, Wang.2026} and application contexts \citep{Kuznetsova.2025}. Experimental work further shows that personalized AI dialogues can substantially reduce conspiracy beliefs \citep{Costello.2024, Boissin.2025}, while awareness of AI involvement does not necessarily diminish the persuasiveness of AI-generated fact-checks \citep{Chae.2024}. Together, these findings suggest that AI assistants hold considerable promise as scalable sensemaking tools.

However, most prior work has examined AI-assisted sensemaking in private, one-on-one settings. Grok's integration into X represents a structurally different form of deployment: an LLM publicly embedded within conversation threads, where AI-generated responses become visible to other users. Recent observational work suggests that information seeking and fact-checking dominate Grok interactions \citep{Mei.2026}, while Grok’s responses align with professional fact-checkers only about \SI{54}{\percent} of the time \citep{Renault.2026}. Yet despite growing interest in AI-assisted sensemaking on social media, little is known about how such systems interact with existing platform-native correction mechanisms.

\textbf{Research gap:} Existing work on LLM-assisted sensemaking has focused on (i) what LLMs are used for in practice \citep{Choudhury.2023, Chatterji.2025, Wang.2024}, (ii) whether AI interactions can shift user beliefs \citep{Costello.2024, Boissin.2025}, and (iii) the fact-checking performance of LLMs \citep{Hoes.2023, Qazi.2026, Quelle.2024, Kuznetsova.2025}. Most of this work has been conducted in controlled settings, leaving the real-world adoption dynamics and the relationship between AI-assisted and platform-native sensemaking poorly understood. We address these gaps through a large-scale multilingual dataset of Grok interactions on X and a systematic examination of their overlap with Community Notes, X's community-based fact-checking system.

\section{Data and Methods}

\subsection{Data Collection}

We collected a random sample of Grok-authored replies via the X API v2 search endpoint between March 7, 2025 -- the date of Grok's integration into X -- and May 28, 2025. To circumvent the API's per-query result cap, we issued search requests in one-minute intervals across the observation window, yielding \num{452045} unique posts in total. For each Grok reply, we retrieved the corresponding prompting post using the reply ID. We then used the prompt post's reply ID to retrieve the post to which the prompt responded, \ie, the prompt target (see Fig.~\ref{fig:data_overview}(a) for an example). This recursive retrieval procedure reconstructs the immediate conversational context of each Grok-user interaction. As illustrated in Fig.~\ref{fig:data_overview}(b), thread structures vary across conversations.  In some cases, the prompt itself constitutes the root post, whereas in others the target post appears further up the thread and is preceded by a separate root post.

Using these data, we constructed structured prompt--reply pairs. We identified Grok replies based on the user ID and linked each reply to its corresponding prompting post using the reply ID. Posts authored by Grok were classified as replies, while the referenced user posts were classified as prompts. We further identified Grok-targeted posts as the initial posts referenced by prompts (\ie, the claims or content to which users directed Grok’s attention). We excluded pairs for which either the prompt or reply could not be retrieved (\eg, due to post deletion or account restrictions), as well as threads initiated by Grok itself and replies generated by other automated accounts (\eg, AskPerplexity \citep{Perplexity.2025}). After these steps, the final sample comprises \num{169137} complete prompt--reply pairs (yielding \num{338274} posts) and \num{69157} distinct Grok-targeted posts, resulting in a total of \num{407431} posts.

\begin{table}[h!]
\centering
\footnotesize
\begin{tabularx}{\columnwidth}{@{}lX@{}}
\toprule
\textbf{Intent} & \textbf{Example Prompts} \\
\midrule
Verification & \textit{``Is this true?''}, \textit{``Is this real?''}, \textit{``Fact-check this''} \\
\addlinespace
Information Request & \textit{``Where does this surname originate from?''}, \textit{``Who is in this picture?''}, \textit{``What is this movie about?''} \\
\addlinespace
Content Understanding & \textit{``Explain this post''}, \textit{``Translate what is said in the video''}, \textit{``Explain this to me like I am a 5th grader''} \\
\addlinespace
Debate & \textit{``What do you think about this?''}, \textit{``Who is right?''}, \textit{``Who's the best football player, Messi or Ronaldo?''} \\
\addlinespace
Humor & \textit{``Generate a funny comment about this''}, \textit{``Based on my posts, roast me''} \\
\addlinespace
Casual Chat & \textit{``What would you name your dog?''}, \textit{``What color do you prefer?''} \\
\addlinespace
Image Generation & \textit{``Ghibli this photo''}, \textit{``Make this car black''}, \textit{``Make it Dragon Ball Z style''} \\
\addlinespace
Harassment & \textit{``Could you make this person bald and obese?''}, \textit{``Define woke for this idiot''}, \textit{``How can you be this dumb?''} \\
\addlinespace
Other & \textit{``Say something''}, \textit{``Why no response?''}, \textit{``Thanks for the answer''} \\
\bottomrule
\end{tabularx}
\caption{Prompt intent categories with illustrative examples drawn from the data.}
\label{tab:intent_examples}
\vspace{-0.5cm}
\end{table}

\subsection{Data Annotation}

We annotated each prompt--reply pair using the \texttt{27B} parameter version of \texttt{Gemma 3} \citep{Gemma.2025} (see SI, Sec.~\nameref{supp:annotation}), given its suitability for processing large volumes of data under computational and cost constraints. Each annotation task included the prompt and the corresponding Grok reply. To provide the model with conversational context, we additionally retrieved one upstream post where available. Contextual posts were selected using a priority rule: we first included the root post of the conversation when available. If no root post could be retrieved, we included the most recent preceding post in the same conversation (see Fig.~\ref{fig:data_overview}(b)). If neither a root post nor a preceding post was present, quoted posts were considered as alternative context. This procedure ensured that each annotation instance contained the immediate conversational context without incorporating the full thread history. The model received multimodal input (text and images; videos were represented using extracted screenshots), enabling classification that incorporated both linguistic and visual information.

The primary construct annotated is communicative intent, \ie the purpose of a Grok invocation. Each prompt was assigned a single intent label from nine categories: \textit{Verification}, \textit{Information Request}, \textit{Content Understanding}, \textit{Debate}, \textit{Humor}, \textit{Casual Chat}, \textit{Image Generation}, \textit{Harassment}, and \textit{Other} (see Tab. ~\ref{tab:intent_examples} for examples). In addition, we annotated both prompts and replies for topical focus using a multi-label scheme covering nine domains: \textit{Politics}, \textit{Economy}, \textit{Health}, \textit{Science}, \textit{Entertainment}, \textit{Society}, \textit{War}, \textit{Crime}, and \textit{Other}. We also annotated the language of both prompts and replies.

\textbf{Validation:}
To validate the LLM-based annotation of intent and topic, we conducted a user study with two research assistants. We drew a stratified sample of 500 prompt–reply pairs based on prompt intent. For each pair, we provided the full conversational context, retrieved any attached media, and translated non-English tweets into English using the same 27B-parameter version of Gemma 3 \citep{Gemma.2025}.
The research assistants were asked to label the intent of the Grok prompt and the topic of the entire thread. Because topics could be multi-labeled, we evaluated performance at the level of individual categories; for consistency, the same approach was applied to intents. LLM-annotator agreement reached an average balanced accuracy of $0.767$ for intent and $0.762$ for topics. Inter-annotator agreement yielded Cohen’s $\kappa$ of $0.673$ for intent and $0.558$ for topics respectively, indicating substantial agreement for intent and moderate agreement for topics.

\section{Empirical Analysis}

We empirically examine how Grok is integrated into platform discourse and how it relates to X's Community Notes system. First, we characterize Grok usage on X (RQ~1) by analyzing invocation contexts, communicative intents, topical domains, and user adoption patterns. Second, we examine the relationship between Grok and Community Notes (RQ~2), focusing on the overlap between the two systems, the types of content that attract both, and their timing.

\subsection{Characterizing Grok Usage on X (RQ~1)}

To understand Grok's role on X, we examine how users invoke the system, who adopts it, and how Grok interactions are embedded within conversation threads. This analysis draws on \num{169137} prompt-reply pairs, corresponding to a total of \num{338274} posts. 

\textbf{Invocation context:} We first examine whether users invoke Grok in direct reply to an existing post or to initialize a new thread (see Fig.~\ref{fig:data_overview}(b)). We find that prompts are predominantly reactive to existing claims or content within ongoing conversations. Specifically, \SI{79.8}{\percent} of prompts are issued as replies within existing threads, while only \SI{20.2}{\percent} occur at the source-post level. Put differently, most Grok interactions are triggered by prior content rather than initiating new discourse. This pattern suggests that Grok operates less as a standalone information tool and more as an instrument embedded in online debate, responding to claims as they emerge.

\begin{figure}[t]
    \centering
    \includegraphics[width=\columnwidth]{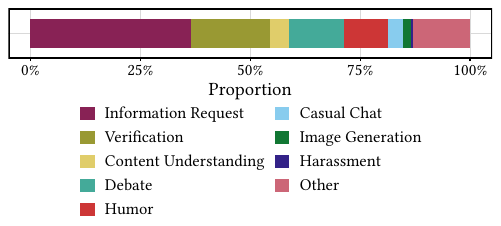}    
    \caption{Distribution of intent categories among Grok prompts.}
    \label{fig:intent_distribution_full_data}
    \vspace{-0.5cm}
\end{figure}

\textbf{Intent:} Grok is predominantly used when users seek additional context or attempt to verify information (see Fig.~\ref{fig:intent_distribution_full_data}). The most frequent intent category is \textit{Information Request} (\SI{36.6}{\percent}), covering general knowledge queries, followed by \textit{Verification} (\SI{17.8}{\percent}), where users ask Grok to assess the truthfulness of a specific claim. Together, these two categories account for \SI{54.4}{\percent} of all prompts, indicating that the majority of Grok interactions fall within fact-checking or contextualization-related use. \textit{Debate} (\SI{12.5}{\percent}) captures prompts where users challenge views or seek supporting arguments, while \textit{Humor} (\SI{9.9}{\percent}) covers jokes and playful content. \textit{Content Understanding} (\SI{4.4}{\percent}) involves requests to explain, summarize, or translate specific provided content -- distinct from general information seeking in that users direct Grok's attention to a concrete artifact rather than a topic. \textit{Harassment}-related prompts are rare (\SI{0.4}{\percent}). Overall, these patterns indicate that Grok is primarily invoked to obtain, clarify, or validate information.

\begin{figure}[t]
    \centering
    \includegraphics[width=\columnwidth]{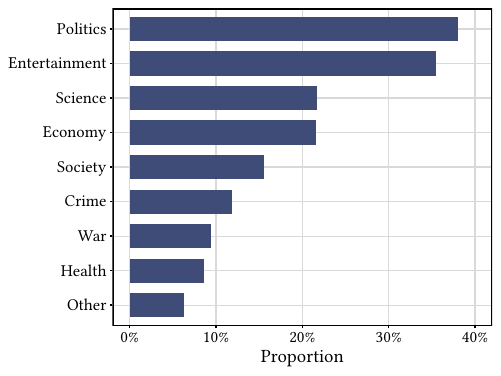}    
    \caption{Distribution of topics among prompt-reply pairs. Because topic labels are multilabel, proportions sum to more than \SI{100}{\percent}.}
    \label{fig:topic_distribution_full_data}
    \vspace{-0.5cm}
\end{figure}

\begin{figure}
    \centering
    \includegraphics[width=\columnwidth]{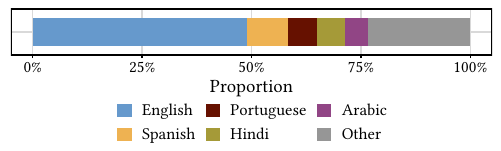}    
    \caption{Language distribution of Grok prompts. English accounts for the largest share, although the majority of prompts are issued in non-English languages.}
    \label{fig:languages_full_data}
    \vspace{-0.5cm}
\end{figure}

\textbf{Topics:} Grok interactions primarily concentrate in politically and information-salient domains (see Fig.~\ref{fig:topic_distribution_full_data}). \textit{Politics} account for \SI{38.0}{\percent} of all topics, followed by \textit{Entertainment} (\SI{35.5}{\percent}), \textit{Science} (\SI{21.8}{\percent}), \textit{Economy} (\SI{21.6}{\percent}), \textit{Society} (\SI{15.6}{\percent}), \textit{Crime} (\SI{11.9}{\percent}), \textit{War} (\SI{9.5}{\percent}), and \textit{Health} (\SI{8.7}{\percent}). Despite the substantial share of \textit{Entertainment}-related interactions, the overall topical distribution suggests Grok is used less as a casual entertainment tool and more as an instrument for navigating complex information. 

\textbf{Languages:} Grok usage is linguistically diverse, indicating broad adoption (see Fig.~\ref{fig:languages_full_data}). \textit{English} accounts for \SI{48.9}{\percent} of all prompts in our data, followed by \textit{Spanish} (\SI{9.3}{\percent}), \textit{Portuguese} (\SI{6.8}{\percent}), \textit{Hindi} (\SI{6.3}{\percent}), and \textit{Arabic} (\SI{5.3}{\percent}), with a heterogeneous residual category accounting for \SI{23.4}{\percent}. Over half of prompts are issued in non-English languages, suggesting that Grok's integration into platform discourse extends beyond English-speaking contexts. 

\textbf{User adoption:}
The majority of users prompt Grok only once. Specifically, \SI{76.8}{\percent} of users issue a single prompt, whereas only \SI{1.46}{\percent} submit five or more prompts. At the same time, the top \SI{1}{\percent} of users account for \SI{8.56}{\percent} of all activity. Together, these patterns suggest that Grok is widely experimented with but rarely used intensively.

Users who invoke Grok multiple times differ marginally from single-prompt users in terms of account characteristics. Multi-prompt users are slightly more likely to hold verified accounts (\SI{13}{\percent} vs. \SI{11}{\percent}) and tend to have more followers and followees. However, KS-tests across user characteristics yield only small effect sizes ($D = [0.03-0.13]$), indicating that repeated use is not concentrated within a structurally distinct user population (see SI, Tab.~\ref{tab:single_multi_user_characteristics}).

More pronounced differences emerge in how users engage with Grok rather than who they are (see SI, Fig.~\ref{fig:intents_topics_by_user_type}(a)). Compared to single-prompt users, multi-prompt users engage more frequently in \textit{Debate} (\SI{18.6}{\percent} vs. \SI{10.3}{\percent}), whereas single-prompt users rely more heavily on \textit{Verification} (\SI{20.2}{\percent} vs. \SI{15}{\percent}) and \textit{Information Request} (\SI{38}{\percent} vs. \SI{34.9}{\percent}). These differences are statistically significant ($\chi^2 = 3470.05$, $p < 0.001$). By contrast, topical distributions remain broadly similar across both groups, although modest but statistically significant differences exist ($\chi^2 = 810.33$, $p < 0.001$): \textit{Entertainment} content is more common among single-prompt users (\SI{37.8}{\percent} vs. \SI{32.6}{\percent}), whereas \textit{Society} topics appear more frequently among multi-prompt users (\SI{17.2}{\percent} vs. \SI{14.2}{\percent}). Overall, these findings suggest that repeated Grok use reflects differences in usage orientation rather than differences in the underlying user population.

\begin{figure}
    \centering
    \includegraphics[width=\columnwidth]{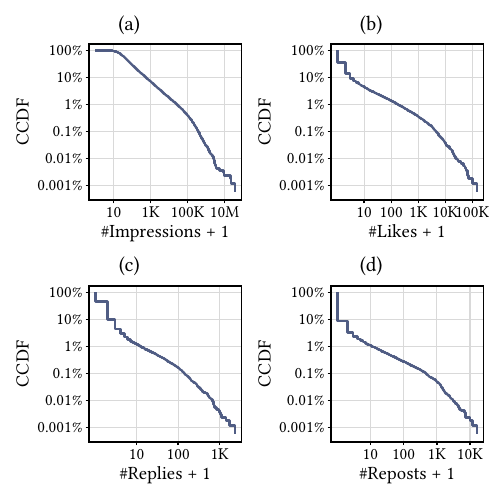}    
    \caption{Complementary cumulative distribution functions (CCDF) of engagement metrics for Grok reply posts ($N$ = 169,137). Panels show the distributions of (a) impressions, (b) likes, (c) replies, and (d) reposts.}
    \label{fig:reply_engagement_panel}
    \vspace{-0.5cm}
\end{figure}

\textbf{Timing \& reach:} Users invoke Grok a median of $5.18$ hours after a conversation thread is initiated, but Grok itself responds rapidly once prompted, with a median response time of $2.68$ minutes. This positions the chatbot as an immediate conversational resource within ongoing discussions. Despite this responsiveness, Grok's replies reach limited audiences. The median reply receives \num{44} impressions and no likes, reposts, or replies (see Fig.~\ref{fig:reply_engagement_panel}), suggesting that Grok-generated information is consumed primarily by the prompting user rather than broadly amplified across the platform.

\begin{figure}[t]
    \centering
    \includegraphics[width=0.93\columnwidth]{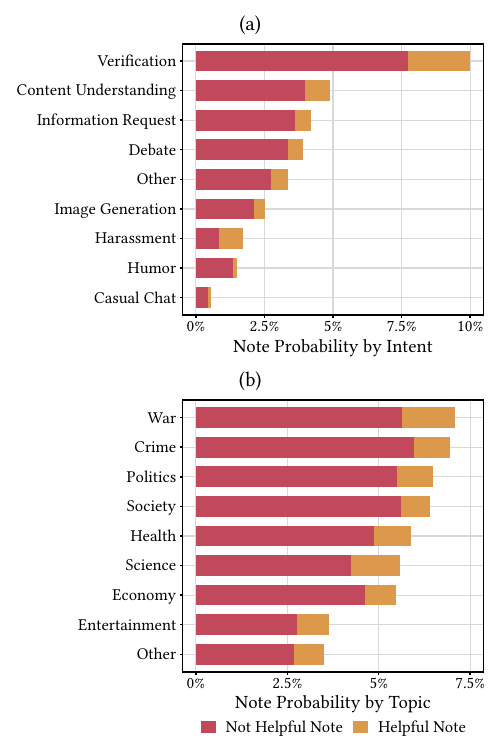}    
    \caption{Probability of receiving a Community Note by (a) prompt intent and (b) topic. Total bar length shows the probability of receiving any note, while the orange segment shows the probability of receiving a helpful note.}
    \label{fig:intents_within_noted_posts}
    \vspace{-0.5cm}
\end{figure}

\begin{figure*}[t]
    \centering
    \includegraphics[width=0.97\textwidth]{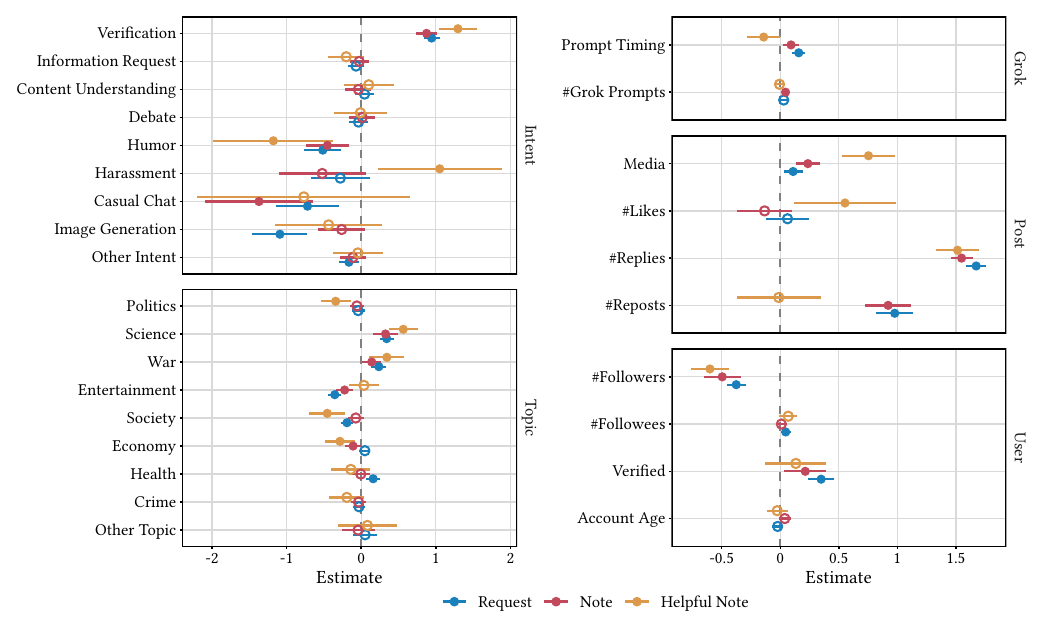}    
    \caption{Logistic regression results of (i) requests, (ii) written notes, and (iii) helpful notes among Grok-targeted posts. Points show log-odds estimates, horizontal lines show \SI{95}{\percent} confidence intervals. Filled points indicate statistical significance ($p < 0.05$), open points indicate non-significance. Standard errors are clustered by target user. $N =$ \num{69157}.}
    \label{fig:coef_plot}
    \vspace{-0.5cm}
\end{figure*}

\subsection{Comparison to Community Notes (RQ~2)}

Given Grok's predominant use for information requests and verification, we now examine how it relates to X's collective fact-checking system, Community Notes, where registered contributors can attach short notes to posts they perceive as misleading or lacking context \citep{Prollochs.2022a, Twitter.2021}. Notes become publicly visible when contributors with diverse rating histories agree they are helpful. In addition, users can submit note requests to signal posts that may require verification \citep{X.2024}. 

Because many Grok prompts involve contextualization and verification, Grok interactions may partially overlap with Community Notes activity. To examine this relationship, we shift the unit of analysis from prompt--reply pairs to the \num{69157} distinct posts targeted by at least one Grok prompt (see Fig.~\ref{fig:data_overview}(b)). We first assess Community Notes prevalence among these posts, then examine which content attracts requests, written notes, and helpful corrections.

\textbf{Community Notes data:} We linked publicly available Community Notes data\footnote{\url{https://communitynotes.x.com/guide/en/under-the-hood/download-data}}, including note metadata, status history, and note request timings, to Grok-targeted posts via their post IDs. Where posts received multiple notes or requests, we aggregated these at the post level, retaining indicators for whether a post received any request or note ($= 1$), the total count of each, and the timing of the first occurrence.

\textbf{Overlap:} Despite Grok's predominant use for verification and contextualization, overlap with Community Notes is limited. Of the \num{69157} posts targeted by Grok prompts, \num{7745}, \ie, \SI{11.2}{\percent}, attracted at least one note request, generating \num{70815} requests in total. Yet only \num{3381} posts (\SI{4.9}{\percent}) received at least one note (see SI, Tab.~\ref{tab:char_by_note_status} for summary statistics). Among the \num{8018} notes written, \SI{94.7}{\percent} classified the Grok-targeted post as potentially misleading, yet only \SI{18.8}{\percent} were ultimately rated helpful and thus made visible. Demand for fact-checking thus outpaces supply, and publicly visible corrections remain comparatively rare. 

As shown in Figure~\ref{fig:intents_within_noted_posts}, the distribution of written and helpful notes among Grok-targeted posts varies substantially across prompt intents and topics. Among prompt intents, \textit{Verification}-targeted posts show the highest probability of receiving both any note (\SI{9.98}{\percent}) and a helpful note (\SI{2.23}{\percent}), followed by \textit{Content Understanding} (\SI{4.88}{\percent}; \SI{0.92}{\percent}) and \textit{Information Request} (\SI{4.20}{\percent}; \SI{0.60}{\percent}) (see Fig.~\ref{fig:intents_within_noted_posts}(a)). By contrast, conversational and \textit{Humor}-related intents rarely attract Community Notes activity, with \textit{Casual Chat} showing the lowest probabilities (\SI{0.57}{\percent} for any note and \SI{0.11}{\percent} for helpful notes). 

A similar, though less pronounced, pattern emerges across topics. Posts related to \textit{War} (\SI{7.09}{\percent}) and \textit{Crime} (\SI{6.94}{\percent}) are most likely to receive notes, followed by \textit{Politics} (\SI{6.49}{\percent}), \textit{Society} (\SI{6.39}{\percent}), \textit{Health} (\SI{5.90}{\percent}), \textit{Science} (\SI{5.59}{\percent}), \textit{Economy} (\SI{5.47}{\percent}). By comparison, \textit{Entertainment} (\SI{3.65}{\percent}) content is substantially less likely to attract corrections (see Fig.~\ref{fig:intents_within_noted_posts}(b)). The distribution of note requests follows a similar pattern and is reported in the Supplementary Information (see SI, Fig.~\ref{fig:intents_within_requested_posts}).

\textbf{Regression model:} To examine which posts are more likely to attract Community Notes activity among Grok-targeted posts, while controlling for confounding factors, we estimate three logistic regression models predicting whether a post receives (i) a note request, (ii) a written note, or (iii) a helpful note. For each post \textit{i} the log-odds of outcome \textit{$Y_i$} is modeled as:

\vspace{-0.1cm}

\begin{equation}
\begin{aligned}
    \text{logit}(Y_i) = \, &\beta_0 
        + \beta_1 \, \text{PromptTiming}_i 
        + \beta_2 \, \text{\#Prompts}_i \\
        &+ \beta_3 \, \textbf{Intent}_i 
        + \beta_4 \, \textbf{Topic}_i 
        + \beta_5 \, \text{Media}_i \\
        &+ \beta_6 \, \text{Likes}_i 
        + \beta_7 \, \text{Replies}_i 
        + \beta_8 \, \text{Reposts}_i \\
        &+ \beta_9 \, \text{Followers}_i 
        + \beta_{10} \, \text{Followees}_i \\
        &+ \beta_{11} \, \text{Verified}_i 
        + \beta_{12} \, \text{AccountAge}_i \\
        &+ \gamma_t + \epsilon_i
\end{aligned}
\end{equation}

\noindent where $Y_i$ denotes one of the three Community Notes outcomes for each post $i$, $Intent_i$ and $Topic_i$ are vectors of dummy indicators, $\gamma_t$ are month-year fixed effects, and $\epsilon_i$ is the error term. We additionally include two Grok-specific predictors. $PromptTiming_i$ captures the log-transformed time elapsed between a post's publication and the first Grok prompt it received, measuring whether early versus late Grok engagement is associated with subsequent Community Notes activity. $\#Prompts_i$ captures the total number of prompts a post received. We further include post-level characteristics ($Media_i$) and engagement metrics ($Likes_i$, $Replies_i$, $Reposts_i$), as well as user-related characteristics such as the number of followers, followees, the account age in years and the verification status. Standard errors are clustered by target user. All continuous predictors are $z$\nobreakdash-standardized, and results are reported as odds ratios and visualized in Figure~\ref{fig:coef_plot}.

\textbf{Coefficient estimates:} The number of Grok prompts directed at a post (\textit{\#Prompts}) is not significantly associated with note requests (coef.: $0.030$, $p = 0.098$) or helpful notes (coef.: $-0.006$, $p = 0.668$), indicating that posts receiving more Grok attention are not substantially more likely to attract correction via the Community Notes system. 

Among intent categories, \textit{Verification} shows the strongest and most consistent associations across all outcomes: verification-targeted posts are $e^{0.944} \approx $ $2.6$ times more likely to receive a note requests (coef.: $0.944$, $p < 0.001$), $2.4$ times more likely to receive any written note (coef.: $0.876$, $p < 0.001$), and $3.7$ times more likely to receive a helpful note (coef.: $1.30$, $p < 0.001$). By contrast, recreational intents such as \textit{Humor}, \textit{Casual Chat} and \textit{Image Generation}, tend to suppress the likelihood of either outcome. 

Across topics, \textit{Science} (coef.: $0.342$ / $0.326$ / $0.564$, all $p < 0.01$) and \textit{War} (coef.: $0.237$ / $0.141$ / $0.344$, all $p < 0.05$) show positive associations across all three outcomes. \textit{Politics} (coef.: $-0.342$, $\text{p} < 0.001$), \textit{Society} (coef.: $-0.455$, $p < 0.001$), and \textit{Economy} (coef.: $-0.285$, $p < 0.01$) are negatively associated with receiving helpful notes.

Post-level predictors play a stronger role in predicting Community Notes activity than user characteristics. The number of replies shows the strongest and most consistent positive associations across outcomes (coef.: $1.67$ / $1.55$ / $1.51$, $p < 0.001$), while reposts strongly predict both requests (coef.: $0.978$, $p < 0.001$) and notes (coef.: $0.921$, $p < 0.001$). Media presence shows a positive association with requests (coef.: $0.110$, $p < 0.01$), notes (coef.: $0.236$, $p < 0.01$), and helpful notes (coef.: $0.753$, $p < 0.001$), whereas follower count is negatively associated with all three outcomes (coef.: $-0.376$ / $-0.495$ / $-0.600$, all $p < 0.001$). Overall, these findings suggest that Community Notes activity is driven more strongly by post visibility and engagement than by account-level characteristics.

\textbf{Model checks:} (1) We assess multicollinearity using variance inflation factors (see SI, Tab.~\ref{tab:vif}), with all predictors remaining below $5$ \citep{Akinwande.2015}. (2) Our results are robust to excluding month-year fixed effects (see SI, Tab.~\ref{tab:regression_nofe}). (3) Since Community Notes skew toward English \citep{Pilarski.2026, Mohammadi.2025} and Grok is primarily trained on English-language data \citep{Boissin.2025}, we re-estimated all models on the subset of English-language prompts and obtained qualitatively similar results (see SI, Tab.~\ref{tab:regression_english}).

\begin{figure}[t]
    \centering
    \includegraphics[width=\columnwidth]{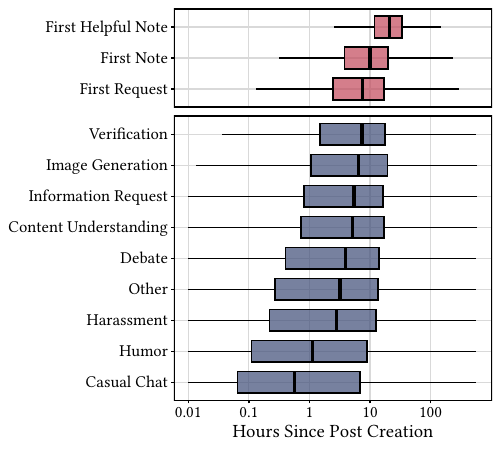}    
    \caption{Timing of Grok prompts by user intent versus request and note timings relative to target post creation.}
    \label{fig:intent_timing_plot}
    \vspace{-0.5cm}
\end{figure}

\textbf{Relative timing:} While the regression analysis identifies which types of content attract activity from both systems, it does not address which responds faster. This is particularly important given criticism that Community Notes operates too slowly to counter misinformation before it spreads \citep{Chuai.2024, Chuai.2024b, Bobek.2026}. 
The timing patterns in our data are consistent with this concern. Across all Grok-targeted posts, the median time from post creation to the first note request is $7.53$ hours, to the first written note $10$ hours, and to the first helpful note $21$ hours. By contrast, Grok responds within minutes once invoked (median: $2.68$ minutes). Consequently, in \SI{85.7}{\percent} of posts that eventually receive a helpful note, Grok is invoked before the correction becomes publicly visible.
However, as Fig.~\ref{fig:intent_timing_plot} illustrates, the delay between post creation and Grok invocation varies by intent: conversational intents such as \textit{Casual Chat} and \textit{Humor} are triggered almost immediately (Medians: $0.12$ and $0.27$ hours respectively), whereas \textit{Verification} prompts arrive latest among all intent categories with a median delay of $6.03$ hours. Nevertheless, even verification-oriented prompts precede the first written note by approximately $3.93$ hours and the first helpful note by $14.93$ hours. These findings suggest that, for most posts that ultimately receive corrections, users encounter AI-generated context well before collective verification becomes available.

\section{Discussion}

\textbf{Relevance:} AI assistants are increasingly integrated into social media platforms \citep{Meta.2024}, but typically operate in private, one-on-one settings. Grok differs in that it can be directly invoked within conversation threads on X, making its responses publicly visible. Unlike expert-based fact-checking or crowd-sourced systems such as Community Notes, which are constrained by scalability and the time required to reach consensus \citep{Chuai.2024, Chuai.2024b, Bobek.2026}, AI-assistants provide immediate, multilingual responses \citep{Hoes.2023,Qazi.2026, Quelle.2024}. Prior work shows that AI-generated content can measurably shift beliefs \citep{Costello.2024, Boissin.2025}, even when users are aware of interacting with an AI assistant \citep{Havin.2025}. Yet little is known about how such systems are used in real-world social media environments  and how they relate to existing sensemaking mechanisms. We contribute to this understanding through a large-scale, multilingual analysis of Grok interactions on X and a systematic examination of how AI-assisted and crowd-based sensemaking coexist and interact on the same platform.

\textbf{Research implications:} Our findings on Grok usage (RQ~1) both align with and extend prior research on AI use. The predominantly reactive nature of Grok invocations positions it as a tool for interpreting existing claims rather than initiating discussion, consistent with theories of sensemaking as a retrospective process \citep{Weick.1995}. The prevalence of information requests and verification-oriented prompts mirrors established patterns of AI use \citep{Chatterji.2025, Choudhury.2023, Wang.2024, Mei.2026}, indicating that embedding AI into social media does not fundamentally alter their core use cases.

At the same time, our findings highlight important differences from prior work. The substantial share of non-English prompts suggests that platform-integrated AI broadens access to information beyond English-speaking contexts \citep{Chatterji.2025}. However, the shallow adoption pattern indicates that this broader reach may not translate into sustained use, qualifying evidence of rapid generative AI adoption \citep{Bick.2024}. This pattern may reflect one-time experimentation or limited awareness of Grok's availability, particularly because invocations tend to be nested within conversation threads rather than prominently surfaced to other users. Moreover, even among users aware of Grok, repeated use may be discouraged due to potential reputational costs associated with publicly relying on AI \citep{Reif.2025}.

Turning to the relationship between Grok and Community Notes (RQ~2), our findings speak to concerns about the speed of crowd-based correction systems \citep{Chuai.2024, Chuai.2024b, Bobek.2026}. Grok responses systematically precede helpful Community Notes, indicating that AI-generated context often becomes available substantially earlier than consensus-based correction. Even verification-oriented prompts, which occur later than other intent categories, precede helpful notes by several hours, suggesting that users routinely encounter AI-generated context before collective corrections. 

However, the two systems differ not only in timing but also in visibility and reach. With a median of only 44 impressions per reply, Grok responses reach few users beyond the original prompter, whereas Community Notes arrive late but are displayed prominently on flagged posts once rated helpful. This limited visibility likely stems from the fact that nearly \SI{80}{\percent} of Grok invocations occur as replies within existing threads, making them less visible to casual readers. As a result, Grok may function more as a private resource than as a broadly visible public intervention, limiting its potential to shape collective sensemaking at scale.

Consistent with this, we find that the number of Grok invocations a post receives has no significant association with subsequent Community Notes activity. This indicates that Community Notes contributors do not systematically react to Grok activity, potentially because Grok prompts and replies remain relatively hidden within conversation threads. Rather than one system triggering the other, the two appear to operate largely in parallel, pointing to a complementary rather than integrated relationship. However, \citet{Renault.2026} report that Grok responses agree with professional fact-checkers only \SI{54.5}{\percent} of the time. Combined with our timing results, this implies that users are often exposed to AI-generated context before any validated correction is available, despite the uncertain accuracy of those responses.

\textbf{Platform implications:} Our findings highlight several opportunities for improving the design of platform-integrated AI assistants. First, because Grok systematically provides context before crowd-based corrections become available, users routinely encounter AI-generated information before any externally validated correction exists. However, this early exposure occurs without external validation, suggesting that platforms should more clearly communicate the uncertainty of AI-generated responses, for example through explicit disclaimers. Second, the persistent imbalance between fact-checking demand and supply \citep{Chuai.2025b, Pilarski.2026} suggests that user-initiated AI prompts could serve as an additional signal for content requiring verification, helping to allocate contributor attention more efficiently. Unlike note requests, which disproportionately target highly visible accounts \citep{Pilarski.2026, Chuai.2025b}, Grok invocations are lower-threshold and more spontaneous, potentially broadening the range of content that receives scrutiny. Third, the observed independence between Grok activity and Community Notes points to opportunities for coordination. Prior work shows that AI can augment crowd-based fact-checking at multiple stages \citep{De.2025, Mohammadi.2025}. Platforms could therefore integrate both systems more closely, for example by enabling AI-generated responses to reference or incorporate existing Community Notes, allowing users to access both immediate context and consensus-based verification within a single interaction.

\textbf{Limitations \& future research: } As with any research, our study has limitations that open avenues for future work. First, we do not directly assess the accuracy of Grok’s responses. Prior work comparing LLM-generated fact-checks on X with professional fact-checkers finds only moderate agreement \citep{Renault.2026}. Future research should therefore examine both the accuracy of AI-generated responses and their alignment with Community Notes. Second, our analysis focuses on a single AI assistant, while others such as AskPerplexity \citep{Perplexity.2025} are also available on X. Whether the patterns observed here generalize across platform-integrated AI assistants remains an open question. Additionally, our relatively short observation window immediately following Grok's introduction means that observed patterns may reflect early-stage experimentation rather than stable adoption. Longitudinal analyses are needed to assess whether users develop sustained reliance on platform-integrated AI over time. Third, while we analyze the co-occurrence of Grok prompts and Community Notes, we cannot determine whether AI-assisted sensemaking influences the likelihood or timing of crowd-based corrections. Finally, we do not capture how users perceive or trust Grok’s responses. This is particularly important given evidence that Community Notes are more trusted than expert fact-checks \citep{Drolsbach.2024} and that AI-generated content can shift beliefs \citep{Costello.2024, Boissin.2025}. Future research should examine how users evaluate AI-generated context relative to crowd-based corrections and how these perceptions shape belief formation.

\section{Conclusion}

This study provides a large-scale, multilingual analysis of how users interact with a publicly embedded AI assistant on social media. Drawing on \num{169137} Grok prompt--reply pairs on X, we examine both how users engage with AI-assisted sensemaking and how these interactions relate to Community Notes. Our findings characterize Grok as a fast, low-threshold layer of public sensemaking that is widely adopted but shallowly used, reactive rather than proactive, and operating in parallel to Community Notes rather than in coordination with it. As AI assistants become increasingly embedded in public discourse, our findings highlight a growing tension between fast, AI-mediated sensemaking and slower, consensus-based verification, underscoring the need to better understand how such systems can be coordinated.

\section{Ethics Statement}
The analyses draw on publicly available data and were conducted in line with ethical practices for research \citep{Rivers.2014}.

\bibliography{references}

\clearpage

\section{Ethics Checklist}

\begin{enumerate}

\item For most authors...
\begin{enumerate}
    \item  Would answering this research question advance science without violating social contracts, such as violating privacy norms, perpetuating unfair profiling, exacerbating the socio-economic divide, or implying disrespect to societies or cultures?
    \answerYes{Yes}
  \item Do your main claims in the abstract and introduction accurately reflect the paper's contributions and scope?
    \answerYes{Yes}
   \item Do you clarify how the proposed methodological approach is appropriate for the claims made? 
    \answerYes{Yes}
   \item Do you clarify what are possible artifacts in the data used, given population-specific distributions?
    \answerYes{Yes}
  \item Did you describe the limitations of your work?
    \answerYes{Yes}
  \item Did you discuss any potential negative societal impacts of your work?
    \answerYes{Yes}
      \item Did you discuss any potential misuse of your work?
    \answerYes{Yes}
    \item Did you describe steps taken to prevent or mitigate potential negative outcomes of the research, such as data and model documentation, data anonymization, responsible release, access control, and the reproducibility of findings?
    \answerYes{Yes}
  \item Have you read the ethics review guidelines and ensured that your paper conforms to them?
    \answerYes{Yes}
\end{enumerate}

\item Additionally, if your study involves hypotheses testing...
\begin{enumerate}
  \item Did you clearly state the assumptions underlying all theoretical results?
    \answerYes{Yes}
  \item Have you provided justifications for all theoretical results?
    \answerYes{Yes}
  \item Did you discuss competing hypotheses or theories that might challenge or complement your theoretical results?
    \answerYes{Yes}
  \item Have you considered alternative mechanisms or explanations that might account for the same outcomes observed in your study?
    \answerYes{Yes}
  \item Did you address potential biases or limitations in your theoretical framework?
    \answerYes{Yes}
  \item Have you related your theoretical results to the existing literature in social science?
    \answerYes{Yes}
  \item Did you discuss the implications of your theoretical results for policy, practice, or further research in the social science domain?
    \answerYes{Yes}
\end{enumerate}

\item Additionally, if you are including theoretical proofs...
\begin{enumerate}
  \item Did you state the full set of assumptions of all theoretical results?
    \answerNA{NA}
	\item Did you include complete proofs of all theoretical results?
    \answerNA{NA}
\end{enumerate}

\item Additionally, if you ran machine learning experiments...
\begin{enumerate}
  \item Did you include the code, data, and instructions needed to reproduce the main experimental results (either in the supplemental material or as a URL)?
    \answerNA{NA}
  \item Did you specify all the training details (e.g., data splits, hyperparameters, how they were chosen)?
    \answerNA{NA}
     \item Did you report error bars (e.g., with respect to the random seed after running experiments multiple times)?
    \answerNA{NA}
	\item Did you include the total amount of compute and the type of resources used (e.g., type of GPUs, internal cluster, or cloud provider)?
    \answerNA{NA}
     \item Do you justify how the proposed evaluation is sufficient and appropriate to the claims made? 
    \answerNA{NA}
     \item Do you discuss what is ``the cost`` of misclassification and fault (in)tolerance?
    \answerNA{NA}
  
\end{enumerate}

\item Additionally, if you are using existing assets (e.g., code, data, models) or curating/releasing new assets, without compromising anonymity...
\begin{enumerate}
  \item If your work uses existing assets, did you cite the creators?
    \answerYes{Yes}
  \item Did you mention the license of the assets?
    \answerNo{No, all datasets are open source and publicly available.}
  \item Did you include any new assets in the supplemental material or as a URL?
    \answerYes{Yes}
  \item Did you discuss whether and how consent was obtained from people whose data you're using/curating?
    \answerYes{Yes}
  \item Did you discuss whether the data you are using/curating contains personally identifiable information or offensive content?
    \answerYes{Yes}
\item If you are curating or releasing new datasets, did you discuss how you intend to make your datasets FAIR?
    \answerNA{NA}
\item If you are curating or releasing new datasets, did you create a Datasheet for the Dataset? 
    \answerNA{NA}
\end{enumerate}

\item Additionally, if you used crowdsourcing or conducted research with human subjects, without compromising anonymity...
\begin{enumerate}
  \item Did you include the full text of instructions given to participants and screenshots?
    \answerNA{NA}
  \item Did you describe any potential participant risks, with mentions of Institutional Review Board (IRB) approvals?
    \answerNA{NA}
  \item Did you include the estimated hourly wage paid to participants and the total amount spent on participant compensation?
    \answerNA{NA}
   \item Did you discuss how data is stored, shared, and deidentified?
    \answerNA{NA}
\end{enumerate}

\end{enumerate}

\clearpage
\appendix
\renewcommand\thetable{S\arabic{table}}
\setcounter{table}{0}
\renewcommand\thefigure{S\arabic{figure}}
\setcounter{figure}{0}

{\LARGE\textbf{Supplementary Materials}}

\section{Annotation}
\label{supp:annotation}

We annotated prompt–reply pairs across a wide range of content characteristics using the \texttt{27B} parameter version of \texttt{Gemma 3}. The corresponding prompt is shown below. 

\smallskip

{\itshape
``You are a research assistant tasked with classifying social media posts based on language, 
topical content, and intent. Return ONLY raw JSON (no markdown, no code fences, no extra text)."

\smallskip
\noindent
Input Sections
\begin{itemize}
    \item Target Post (optional): Context only. Do NOT label this section.
    \item Prompt Post (required): The user's original message or question. May be unavailable if deleted.
    \item Reply Post (required): The assistant's response.
\end{itemize}

\noindent
Labeling Rules
\begin{itemize}
    \item Use the Target Post only as context; never include it in the output.
    \item Prompt Post: label `language`, 
    `intent`, and `topics`.
    \item Reply Post: label `language`, 
    and `topics`.
\end{itemize}

\noindent
Definitions and Output Format:

\begin{enumerate}

\item Language

 ISO 639-1 code (e.g., `en`, `de`, `fr`, `es`). If unclear, return `[unknown]`.

\item Intent

Exactly one must be `true`; all others `false`.
\begin{itemize}
    \item Verification: asks if a claim/story is true, wants fact-check/evidence/sources.
    \item Information Request: asking for general knowledge/information about topics (what is X? who is Y? how does Z work?).
    \item Content Understanding: help understanding/working with specific provided content (explain THIS, what does THIS mean, translate THIS, summarize THIS).
    \item Debate: challenging views, arguing positions, engaging in debate.
    \item Humor: jokes, memes, playful content.
    \item Harassment: hostile, abusive, or threatening language.
    \item Image Generation: create or modify an image.
    \item Casual Chat: greetings, small talk, casual conversation.
    \item Other: only if prompt missing or none fit.
\end{itemize}

\item Topic

Set `true` only if the topic is a substantial focus of the post. If no other topic is `true`, set `Other` $=$ `true`. Possible topics are `Politics`, `Entertainment`, `Science`, `Economy`, `Society`, `Health`.

\end{enumerate}
}

\section{Summary Statistics}
\label{supp:summary}

\begin{figure}[t]
    \centering
    \includegraphics[width=\columnwidth]{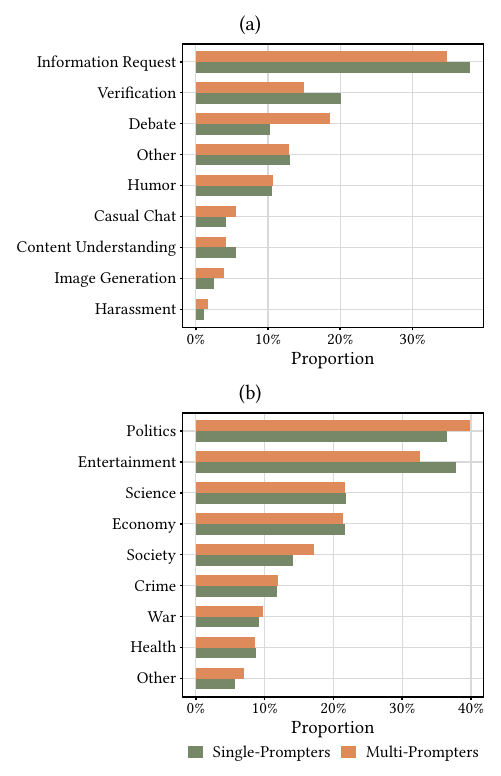}
    \caption{Distribution of prompt intents and topics by user type (single- vs. multi-prompt users).}
    \label{fig:intents_topics_by_user_type}
    \vspace{-0.5cm}
\end{figure}

\begin{figure}[t]
    \centering
    \includegraphics[width=\columnwidth]{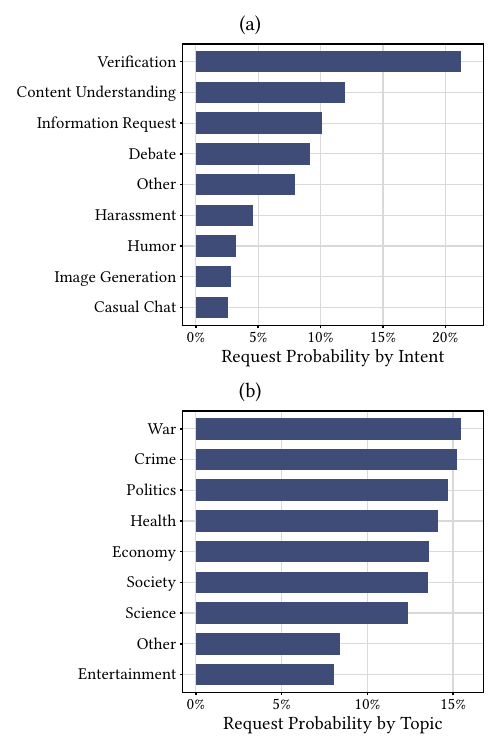}    
    \caption{Probability of receiving a Community Note request by (a) prompt intent and (b) topic.}
    \label{fig:intents_within_requested_posts}
    \vspace{-0.5cm}
\end{figure}

This section reports descriptive statistics supplementing the main analysis. Table~\ref{tab:single_multi_user_characteristics} and Figure~\ref{fig:intents_topics_by_user_type} characterize users and usage patterns by prompt type, Figure~\ref{fig:intents_within_requested_posts} reports note request probabilities by intent and topic, and Table~\ref{tab:char_by_note_status} summarizes post and user characteristics by Community Notes status. Table~\ref{tab:vif} reports VIF for the regression models.

\begin{table}[h]
\centering
\footnotesize
\begin{tabularx}{\columnwidth}{@{\extracolsep{\fill}}lcc}
\toprule
 & (1) & (2) \\
 & Single-Prompt & Multi-Prompt \\
\midrule
\#Users & 93,579 & 28,214 \\
[.5mm]
\textit{User Characteristics} & & \\[-2.5mm]
\rule{3cm}{0.1pt} \\
Avg. Prompts & 1.00 & 2.68 \\
Verified (\%) & 11.0\% & 13.0\% \\
Median Followers & 2,965 & 3,427 \\
Median Followees & 869 & 1,090 \\
Median Account Age (Years) & 6.64 & 6.36 \\
[.5mm]
\textit{Post Characteristics} & & \\[-2.5mm]
\rule{3cm}{0.1pt} \\
Median Likes & 18.46 & 22.29 \\
Median Replies & 2.21 & 2.18 \\
Median Reposts & 1.42 & 2.16 \\
Median Impressions & 4,564 & 3,443 \\
\bottomrule
\end{tabularx}
\caption{Characteristics by user type. Continuous features are reported as medians; \textit{Verified} is reported as a share.}
\label{tab:single_multi_user_characteristics}
\end{table}

\begin{table}[h!]
\centering
\small
\begin{tabular}{lccc}
\toprule
 & (1) & (2) & (3) \\
Variable & Request & Note & Helpful \\
\midrule
PromptTiming             & 1.123 & 1.126 & 1.126 \\
\#Prompts                & 1.277 & 1.319 & 1.132 \\
\textit{Intent:} Verification     & 1.789 & 1.665 & 1.481 \\
\textit{Intent:} Information Request   & 1.666 & 1.527 & 1.320 \\
\textit{Intent:} Content Understanding & 1.180 & 1.130 & 1.093 \\
\textit{Intent:} Debate           & 1.229 & 1.176 & 1.111 \\
\textit{Intent:} Humor            & 1.093 & 1.099 & 1.060 \\
\textit{Intent:} Casual Chat      & 1.060 & 1.079 & 1.013 \\
\textit{Intent:} Image Generation       & 1.066 & 1.106 & 1.066 \\
\textit{Intent:} Harassment      & 1.018 & 1.017 & 1.059 \\
\textit{Intent:} Other           & 1.252 & 1.188 & 1.126 \\
\textit{Topic:} Politics          & 1.241 & 1.235 & 1.253 \\
\textit{Topic:} Economy           & 1.080 & 1.079 & 1.069 \\
\textit{Topic:} Health            & 1.037 & 1.037 & 1.036 \\
\textit{Topic:} Science           & 1.106 & 1.106 & 1.115 \\
\textit{Topic:} Entertainment     & 1.192 & 1.171 & 1.176 \\
\textit{Topic:} Society          & 1.069 & 1.069 & 1.061 \\
\textit{Topic:} War               & 1.088 & 1.088 & 1.115 \\
\textit{Topic:} Crime             & 1.061 & 1.065 & 1.059 \\
\textit{Topic:} Other             & 1.079 & 1.072 & 1.077 \\
Media                   & 1.036 & 1.035 & 1.031 \\
Likes                    & 2.562 & 2.786 & 2.905 \\
Replies                 & 1.401 & 1.513 & 1.609 \\
Reposts                  & 2.425 & 2.678 & 2.830 \\
Followers                & 1.400 & 1.451 & 1.451 \\
Followees                & 1.030 & 1.035 & 1.037 \\
Verified                & 1.163 & 1.175 & 1.203 \\
AccountAge               & 1.123 & 1.130 & 1.109 \\
\bottomrule
\end{tabular}
\caption{VIF for each logistic regression model predicting (1) requests, (2) notes, and (3) helpful notes.}
\label{tab:vif}
\end{table}

\begin{table*}
\centering
\footnotesize
\begin{tabularx}{0.9\textwidth}{@{\extracolsep{\fill}}lccc}
\toprule
 & (1) & (2) & (3) \\
 & No Note & Any Note & Helpful Note \\
\midrule
\#Posts & \num{65776} & \num{3381} & \num{635} \\
[.5mm]
\textit{Post Characteristics} & & & \\[-2.5mm]
\rule{3cm}{0.1pt} \\
Impressions & \num{956701} (\num{5343921}) & \num{6893180} (\num{22046651}) & \num{4233116} (\num{7226378}) \\
Likes & \num{8734} (\num{33310}) & \num{49058} (\num{88187}) & \num{34466} (\num{50742}) \\
Replies & \num{393} (\num{1668}) & \num{3136} (\num{6819}) & \num{1844} (\num{2559}) \\
Reposts & \num{983} (\num{3522}) & \num{6915} (\num{14017}) & \num{4093} (\num{5593}) \\
Quotes & \num{116} (\num{589}) & \num{863} (\num{3293}) & \num{698} (\num{1230}) \\
Media & \SI{49.0}{\percent} & \SI{73.3}{\percent} & \SI{84.1}{\percent} \\
[.5mm]
\textit{User Characteristics} & & & \\[-2.5mm]
\rule{3cm}{0.1pt} \\
Account Age (Years) & \num{7.11} (\num{5.20}) & \num{8.19} (\num{5.41}) & \num{7.29} (\num{5.15}) \\
Verified & \SI{58.9}{\percent} & \SI{87.6}{\percent} & \SI{83.5}{\percent} \\
\#Followers & \num{2037252} (\num{16829588}) & \num{7994937} (\num{37009855}) & \num{1263628} (\num{9051420}) \\
\#Followees & \num{4096} (\num{21673}) & \num{8534} (\num{35321}) & \num{9162} (\num{24870}) \\
\bottomrule
\end{tabularx}
\caption{Post and user characteristics by note status. Binary features are reported as shares, while continuous features are reported as means with standard deviations in parentheses.}
\label{tab:char_by_note_status}    
\vspace{-0.5cm}
\end{table*}

\clearpage
\onecolumn

\section{Robustness Checks}
\label{supp:robustness}

Tables~\ref{tab:regression_nofe} and~\ref{tab:regression_english} report two robustness checks for the main regression model (Figure~\ref{fig:coef_plot}). Table~\ref{tab:regression_nofe} re-estimates all three models excluding month-year fixed effects. Table~\ref{tab:regression_english} restricts the sample to posts targeted by at least one English-language prompt, addressing potential confounds arising from the skew of Community Notes participation toward English-speaking contributors \citep{Pilarski.2026, Mohammadi.2025}.

\begin{table}[h]
\centering
\small
\begin{tabular}{l
S[table-format=3.3, table-space-text-post={***}]
S[table-format=1.3]
S[table-format=3.3, table-space-text-post={***}]
S[table-format=1.3]
S[table-format=3.3, table-space-text-post={***}]
S[table-format=1.3]
}
\toprule
 & \multicolumn{2}{c}{(1)} & \multicolumn{2}{c}{(2)} & \multicolumn{2}{c}{(3)} \\
 & \multicolumn{2}{c}{Request} & \multicolumn{2}{c}{Note} & \multicolumn{2}{c}{Helpful Note} \\
\cmidrule(lr){2-3} \cmidrule(lr){4-5} \cmidrule(lr){6-7}
 & {Coef.} & {Std. Error} & {Coef.} & {Std. Error} & {Coef.} & {Std. Error} \\
\midrule
\textit{Grok} & & & & & & \\[-2.5mm]
\rule{3cm}{0.1pt} \\
Prompt Timing  & 0.175{$^{***}$} & (0.025) & 0.065{}         & (0.033) & -0.149{$^{*}$}  & (0.067) \\
\#Grok Prompts & 0.032{}         & (0.018) & 0.043{$^{**}$}  & (0.017) & -0.006{}        & (0.014) \\
\addlinespace[4pt]
\textit{Intent} & & & & & & \\[-2.5mm]
\rule{3cm}{0.1pt} \\
Verification          & 0.933{$^{***}$}  & (0.054) & 0.876{$^{***}$}  & (0.072) & 1.310{$^{***}$} & (0.129) \\
Information Request   & -0.050{}         & (0.052) & -0.029{}         & (0.066) & -0.188{}        & (0.124) \\
Content Understanding & 0.057{}          & (0.067) & -0.039{}         & (0.089) & 0.111{}         & (0.171) \\
Debate                & -0.002{}         & (0.063) & 0.016{}          & (0.089) & 0.009{}         & (0.183) \\
Humor                 & -0.485{$^{***}$} & (0.125) & -0.446{$^{***}$} & (0.146) & -1.170{$^{**}$} & (0.410) \\
Harassment            & -0.296{}         & (0.207) & -0.521{}         & (0.298) & 1.030{$^{*}$}   & (0.424) \\
Casual Chat           & -0.604{$^{**}$}  & (0.211) & -1.320{$^{***}$} & (0.377) & -0.722{}        & (0.726) \\
Image Generation      & -1.370{$^{***}$} & (0.195) & -0.293{}         & (0.161) & -0.520{}        & (0.363) \\
Other                 & -0.128{}         & (0.068) & -0.113{}         & (0.091) & -0.021{}        & (0.171) \\
\addlinespace[4pt]
\textit{Topic} & & & & & & \\[-2.5mm]
\rule{3cm}{0.1pt} \\
Politics      & -0.024{}        & (0.045) & -0.057{}         & (0.050) & -0.346{$^{***}$} & (0.103) \\
Economy       & 0.076{$^{*}$}   & (0.037) & -0.104{}         & (0.054) & -0.275{$^{**}$}  & (0.104) \\
Health        & 0.135{$^{**}$}  & (0.049) & -0.003{}         & (0.066) & -0.146{}         & (0.133) \\
Science       & 0.337{$^{***}$} & (0.049) & 0.324{$^{***}$}  & (0.086) & 0.556{$^{***}$}  & (0.098) \\
Entertainment & -0.357{$^{***}$}& (0.044) & -0.223{$^{***}$} & (0.056) & 0.025{}          & (0.105) \\
Society       & -0.162{$^{***}$}& (0.043) & -0.068{}         & (0.053) & -0.450{$^{***}$} & (0.123) \\
War           & 0.140{$^{**}$}  & (0.051) & 0.135{$^{*}$}    & (0.064) & 0.310{$^{**}$}   & (0.120) \\
Crime         & -0.035{}        & (0.043) & -0.036{}         & (0.054) & -0.198{}         & (0.120) \\
Other         & 0.061{}         & (0.079) & -0.039{}         & (0.113) & 0.067{}          & (0.199) \\
\addlinespace[4pt]
\textit{Post} & & & & & & \\[-2.5mm]
\rule{3cm}{0.1pt} \\
Media      & 0.104{$^{**}$}  & (0.041) & 0.240{$^{***}$} & (0.051) & 0.758{$^{***}$} & (0.114) \\
\#Likes    & 0.074{}         & (0.092) & -0.112{}        & (0.120) & 0.560{$^{*}$}   & (0.220) \\
\#Replies  & 1.660{$^{***}$} & (0.042) & 1.550{$^{***}$} & (0.048) & 1.510{$^{***}$} & (0.093) \\
\#Reposts  & 0.944{$^{***}$} & (0.079) & 0.904{$^{***}$} & (0.101) & -0.030{}        & (0.182) \\
\addlinespace[4pt]
\textit{User} & & & & & & \\[-2.5mm]
\rule{3cm}{0.1pt} \\
\#Followers  & -0.366{$^{***}$} & (0.041) & -0.501{$^{***}$} & (0.079) & -0.598{$^{***}$} & (0.082) \\
\#Followees  & 0.042{$^{*}$}    & (0.021) & 0.012{}          & (0.023) & 0.070{}          & (0.038) \\
Verified     & 0.332{$^{***}$}  & (0.056) & 0.215{$^{*}$}    & (0.091) & 0.119{}          & (0.132) \\
Account Age  & -0.022{}         & (0.023) & 0.039{}          & (0.027) & -0.027{}         & (0.046) \\
\midrule
Month-Year FE  & \multicolumn{2}{c}{No} & \multicolumn{2}{c}{No} & \multicolumn{2}{c}{No} \\
Clustered SE   & \multicolumn{2}{c}{Target User} & \multicolumn{2}{c}{Target User} & \multicolumn{2}{c}{Target User} \\
\#Observations & \multicolumn{2}{c}{69,157} & \multicolumn{2}{c}{69,157} & \multicolumn{2}{c}{69,157} \\
\bottomrule
\multicolumn{7}{l}{\scriptsize{$^{***}p<0.001$; $^{**}p<0.01$; $^{*}p<0.05$}} \\
\end{tabular}
\caption{Logistic regression results excluding month-year fixed effects. Dependent variables are whether a post received (1) a note request, (2) a written note, or (3) a helpful note. Standard errors are clustered by target user. All continuous predictors are $z$\nobreakdash-standardized.}
\label{tab:regression_nofe}
\end{table}

\begin{table*}[t]
\small
\centering
\begin{tabular}{l
S[table-format=3.3, table-space-text-post={***}]
S[table-format=1.3]
S[table-format=3.3, table-space-text-post={***}]
S[table-format=1.3]
S[table-format=3.3, table-space-text-post={***}]
S[table-format=1.3]
}
\toprule
 & \multicolumn{2}{c}{(1)} & \multicolumn{2}{c}{(2)} & \multicolumn{2}{c}{(3)} \\
 & \multicolumn{2}{c}{Request} & \multicolumn{2}{c}{Note} & \multicolumn{2}{c}{Helpful Note} \\
\cmidrule(lr){2-3} \cmidrule(lr){4-5} \cmidrule(lr){6-7}
 & {Coef.} & {Std. Error} & {Coef.} & {Std. Error} & {Coef.} & {Std. Error} \\
 \midrule
\textit{Grok} & & & & & & \\[-2.5mm]
\rule{3cm}{0.1pt} \\
Prompt Timing  & 0.165{$^{***}$} & (0.034) & 0.092{$^{*}$}   & (0.044) & -0.149{}        & (0.089) \\
\#Grok Prompts & 0.037{}         & (0.021) & 0.059{$^{**}$}  & (0.021) & -0.007{}        & (0.014) \\
\addlinespace[4pt]
\textit{Intent} & & & & & & \\[-2.5mm]
\rule{3cm}{0.1pt} \\
Verification          & 0.935{$^{***}$}  & (0.066) & 0.857{$^{***}$}  & (0.083) & 1.330{$^{***}$} & (0.163) \\
Information Request   & -0.087{}         & (0.064) & -0.032{}         & (0.078) & -0.268{}        & (0.154) \\
Content Understanding & -0.025{}         & (0.083) & -0.006{}         & (0.103) & 0.200{}         & (0.203) \\
Debate                & -0.027{}         & (0.082) & -0.059{}         & (0.105) & -0.122{}        & (0.249) \\
Humor                 & -0.541{$^{***}$} & (0.157) & -0.292{}         & (0.173) & -0.971{$^{*}$}  & (0.493) \\
Harassment            & -0.220{}         & (0.226) & -0.444{}         & (0.323) & 1.350{$^{**}$}  & (0.482) \\
Casual Chat           & -0.743{$^{*}$}   & (0.303) & -2.090{$^{***}$} & (0.650) & -0.520{}        & (1.020) \\
Image Generation      & -1.090{$^{***}$} & (0.215) & -0.368{$^{*}$}   & (0.184) & -1.060{$^{*}$}  & (0.535) \\
Other                 & -0.238{$^{**}$}  & (0.089) & -0.100{}         & (0.113) & -0.183{}        & (0.233) \\
\addlinespace[4pt]
\textit{Topic} & & & & & & \\[-2.5mm]
\rule{3cm}{0.1pt} \\
Politics      & 0.085{}         & (0.059) & 0.006{}          & (0.061) & -0.218{}         & (0.129) \\
Economy       & -0.033{}        & (0.046) & -0.163{$^{**}$}  & (0.062) & -0.422{$^{***}$} & (0.132) \\
Health        & 0.194{$^{***}$} & (0.060) & -0.063{}         & (0.079) & -0.082{}         & (0.162) \\
Science       & 0.221{$^{***}$} & (0.056) & 0.321{$^{***}$}  & (0.094) & 0.544{$^{***}$}  & (0.120) \\
Entertainment & -0.380{$^{***}$}& (0.056) & -0.240{$^{***}$} & (0.065) & 0.186{}          & (0.128) \\
Society       & -0.276{$^{***}$}& (0.055) & -0.180{$^{**}$}  & (0.064) & -0.711{$^{***}$} & (0.160) \\
War           & 0.325{$^{***}$} & (0.063) & 0.239{$^{***}$}  & (0.075) & 0.469{$^{***}$}  & (0.142) \\
Crime         & 0.007{}         & (0.055) & 0.017{}          & (0.064) & -0.240{}         & (0.149) \\
Other         & 0.030{}         & (0.115) & -0.053{}         & (0.156) & 0.162{}          & (0.283) \\
\addlinespace[4pt]
\textit{Post} & & & & & & \\[-2.5mm]
\rule{3cm}{0.1pt} \\
Media      & 0.185{$^{***}$} & (0.050) & 0.262{$^{***}$} & (0.064) & 0.790{$^{***}$} & (0.146) \\
\#Likes    & 0.127{}         & (0.124) & -0.029{}        & (0.157) & 0.516{}         & (0.294) \\
\#Replies  & 1.780{$^{***}$} & (0.057) & 1.590{$^{***}$} & (0.060) & 1.570{$^{***}$} & (0.118) \\
\#Reposts  & 0.768{$^{***}$} & (0.108) & 0.832{$^{***}$} & (0.132) & -0.019{}        & (0.252) \\
\addlinespace[4pt]
\textit{User} & & & & & & \\[-2.5mm]
\rule{3cm}{0.1pt} \\
\#Followers  & -0.436{$^{***}$} & (0.049) & -0.596{$^{***}$} & (0.079) & -0.669{$^{***}$} & (0.104) \\
\#Followees  & 0.041{}          & (0.027) & -0.010{}         & (0.027) & 0.037{}          & (0.047) \\
Verified     & 0.283{$^{***}$}  & (0.078) & 0.241{$^{*}$}    & (0.106) & 0.124{}          & (0.182) \\
Account Age  & -0.013{}         & (0.028) & 0.013{}          & (0.032) & -0.112{}         & (0.058) \\
\midrule
Month-Year FE  & \multicolumn{2}{c}{Yes} & \multicolumn{2}{c}{Yes} & \multicolumn{2}{c}{Yes} \\
Clustered SE   & \multicolumn{2}{c}{Target User} & \multicolumn{2}{c}{Target User} & \multicolumn{2}{c}{Target User} \\
\#Observations & \multicolumn{2}{c}{60,744} & \multicolumn{2}{c}{60,744} & \multicolumn{2}{c}{60,744} \\
\bottomrule
\multicolumn{7}{l}{\scriptsize{$^{***}p<0.001$; $^{**}p<0.01$; $^{*}p<0.05$}} \\
\end{tabular}
\caption{Logistic regression results restricting the sample to posts targeted by at least one English-language prompt. Dependent variables are whether a post received (1) a note request, (2) a written note, or (3) a helpful note. Month-year fixed effects and standard errors clustered by target user are included in all models. All continuous predictors are $z$\nobreakdash-standardized.}
\label{tab:regression_english}
\end{table*}

\end{document}